\documentclass[12pt]{amsart}
\usepackage{amssymb,amsmath,graphics,graphicx}
\usepackage{color}

\hoffset - 0.9 truecm
\def\R{{\mathbb{R}}}

\def\cP{{\mathcal P}}
\def\bE{{\bf E}}
\def\bq{{\bf q}}
\def\bm{{\bf m}}
\def\bj{{\bf j}}
\def\bk{{\bf k}}
\def\bD{{\bf D}}
\def\bB{{\bf B}}
\def\bH{{\bf H}}
\def\bp{{\bf p}}
\def\bM{{\bf M}}
\def\bJ{{\bf J}}
\def\k{\kappa}
\def\s{\sigma}
\def\t{\theta}

\def\W{\Omega}

\def\e{\varepsilon}

\def\div{\nabla \cdot}
\def\rot{\nabla \times}

\def\vf{\varphi}

\begin{document}
\title{A non isothermal phase-field model for the ferromagnetic transition}
\author{V. Berti}
\address[V. Berti]{
University of Bologna\\ Department of Mathematics\\
Piazza di Porta S. Donato 5\\ 40126 Bologna\\ Italy.
} 
\email{berti@dm.unibo.it}

\author{D. Grandi}
\address[D. Grandi]{
University of Bologna\\ Department of Mathematics\\
Piazza di Porta S. Donato 5\\ 40126 Bologna\\ Italy.
} 
\email{grandi@dm.unibo.it}

\baselineskip=18pt

\begin{abstract}
We propose a model for non isothermal ferromagnetic phase transition based on a phase field approach, in which the phase parameter is related but not identified with the magnetization. The magnetization is split in a paramagnetic and in a ferromagnetic contribution, dependent on a scalar phase parameter and identically null above the Curie temperature. The dynamics of the magnetization below the Curie temperature is governed by the order parameter evolution equation and by a Landau-Lifshitz type equation for the magnetization vector. In the simple situation of a uniaxial magnet it is shown how the order parameter dynamics reproduces the hysteresis effect of the magnetization.
\end{abstract}

\maketitle

\section{Introduction}

The peculiar feature of \emph{ferromagnetic} materials is the behaviour of the magnetization vector below a characteristic value $\theta_c$ of temperature, named the \emph{Curie temperature}. Firstly, at temperatures $\theta<\theta_c$, a non zero value $\bM_0(\theta)$ of the magnetization is stable even at zero external magnetic fields; this magnetization is said the \emph{spontaneous magnetization}. On the contrary, in the \emph{paramagnetic} regime, that is for $\theta>\theta_c$, the magnetization vanishes at zero external field. Moreover, the way in which external fields influence the magnetization vector differs in the two cases. While in the paramagnetic state the magnetization at a point is a function of the magnetic fields at that point (with a proportionality relation at sufficiently low fields), in the ferromagnetic state the magnetization shows well known hysteresis phenomena \cite{BS}. In the ferromagnetic materials the external fields influence primarily the direc
 tion of the magnetization vector, and a model for the magnetization evolution in saturation conditions has been proposed long ago by Landau and Lifshitz \cite{LL1935,LL}.\\
These qualitative differences in the behaviour of the magnetization on the two sides of the Curie temperature can be understood in the framework of the phase transitions. The clarification of this issue is a fundamental contribution of the \emph{Landau theory} of phase transitions \cite{LL-stat-mech,G}. The approach of Landau is based on the concept of an \emph{order parameter}, which is a physical (macroscopical) observable quantity whose behaviour is able to account for the microscopical change of structure which generally characterizes phase transitions. Strictly speaking, the Landau theory accounts for second order (or continuous) phase transitions, which, according to Landau, are properly understood in terms of \emph{symmetry breaking}. So,  in the case of ferromagnetic transition, for example, the order parameter of the Landau theory is the magnetization vector: the transition manifests itself as rotational symmetry breaking due to the set up of a non-zero magnetization
 
 $\bM_0$ under otherwise isotropic conditions, that is in zero external field. Perhaps, it is not superfluous to point out that the value of the order parameter is not in itself an indicator of the phase of the material, except in the particular case of null external field.\\
Along these lines, in \cite{BFG} a three dimensional evolutive model is proposed, using the whole magnetization vector with the order parameter, ruled by a vectorial Ginzburg-Landau time-dependent equation.\\
In this paper we propose a model different from the original Landau setting and closer to a general phase field approach. That is, we introduce a \emph{scalar} phase field, which (unlike the magnetization vector) vanishes above the Curie temperature,
even in presence of external fields. The order parameter vanishes in a continuous way when the Curie temperature is approached, according to the second order character of the transition. The magnetization will be decomposed in two contributions, one which is of paramagnetic character, namely it is a direct function of the external field, and an other one, depending also on the order parameter, which is considered as an independent field with its own evolution equation (the time-dependent Ginzburg-Landau equation). The sense of this separation is not, of course, that of a physical distinction between two different sources of magnetization, as well as, for example, the two fluid theory of superfluidity \cite{Tilley} is not the theory of a mixture of fluids. The two magnetization contributions reflect the different way in which the magnetization evolves under the external field in the ferromagnetic and in the paramagnetic regime. In particular, the hysteresis phenomena manifesti
 ng themselves in the ferromagnetic phase are determined by the coupling with the phase field evolution equation. Under this respect, the model draws on the internal variable models, in which a  history-dependent constitutive equation for a physical quantity (in our case the magnetization) is obtained through the coupling with an internal variable obeying its own differential evolution equation (\cite{Vis}). 
Nevertheless, in this model, the phase field is not a mere internal variable, nor its evolution equation is a constitutive one. We assume that the phase field equation is a true balance equation associated with its own power balance \cite{Fab}.\\
 The phase field of this model is related to the modulus of the spontaneous magnetization (nevertheless it is influenced by the external field); loosely speaking, we can say that its physical meaning is related to the microscopic order set up by the microscopic exchange interactions, responsible of spontaneous magnetization.\\
The model we are proposing couples the Maxwell equations for the electromagnetic field with a scalar time-dependent Ginzburg-Landau equation for the order parameter and a heat balance equation for the temperature. Two constitutive equations define the relation between the magnetization and the order parameter: the magnetization is split in a paramagnetic and a ferromagnetic contribution and the direction of the last one, in the tridimensional case, is ruled by a Landau-Lifshitz-Gilbert equation \cite{LL1935, Gilbert}, 
suitably modified in order to describe the non saturated regime. The constitutive choices of the model are verified to be consistent with the second law of thermodynamics in the form of the Clausius-Duhem inequality.\\
Finally, we present a particular case of the model when the direction of the magnetization is fixed, as in the case of uniaxial ferromagnets (one-dimensional model). We show in this situation how the dynamics of the order parameter in the ferromagnetic phase gives the usual shape of the hysteresis cycle of the magnetization. 

\section{Three dimensional model}
Let us consider a ferromagnetic material occupying a bounded domain $\W\subset\R^3$. 
Denoting by $\bE, \bH, \bD, \bB$ the electric field, the magnetic field, the electric displacement and the magnetic induction, the behavior of the material is ruled by Maxwell's equations
\begin{eqnarray}
\label{Max1}
\rot\bE&=&-\dot\bB,\qquad\qquad \rot\bH=\dot\bD+\bJ,\\
\label{Max2}
\div\bB&=&0,\qquad\qquad\quad\ \ \div\bD=\rho,	
\end{eqnarray}
where $\bJ$ is the current density and $\rho$ is the charge density. We assume the constitutive equations
\begin{equation}	\label{eqcost}
\bD=\e\bE,\qquad\qquad \bB=\mu_0\bH+\bM,\qquad\qquad\bJ=\s\bE,
\end{equation}
where $\e,\mu_0,\s$ are respectively the dielectric constant, the magnetic permeability and the conductivity, while $\bM$ denotes the magnetization.
As known, in paramagnetic materials, the magnetization is a function of the magnetic field $\bH$. On the contrary, ferromagnetic systems
are characterized by a time-non local relation between magnetization and the magnetic field. Therefore, in order to describe paramagnetic-ferromagnetic transitions, we write
\begin{equation}\label{B}
\bB=\mu\bH+\hat{\bM}, \qquad \mu=\mu(\bH,\t).
\end{equation}
This amounts to split the magnetization as $\bM=(\mu-\mu_0)\bH+\hat{\bM}$, 
where $\hat\bM$ is the part of the magnetization whose value at a given time cannot be expressed as function of the field $\bH$ at the same time. This contribution exists only in the ferromagnetic phase and it is history-dependent, so $\hat\bM=0$ in the paramagnetic state and $\hat \bM\neq 0$ in the ferromagnetic regime. 
The model we propose is set in the general context of the Ginzburg-Landau theory by defining, as order parameter, a scalar phase variable $\vf$ such that $\vf=0$ in the paramagnetic phase and  $\vf> 0$ in the ferromagnetic state.
As a consequence the magnetization is related to $\vf$.
More precisely, we assume that   
\begin{equation}\label{M}
\hat\bM=M(\vf,\t)\bm, \qquad M(\vf,\t)\geq 0
\end{equation}
where $\bm$ is a unit versor and the modulus $M(\vf,\t)$ depends also on the temperature.

The evolution of the phase $\vf$ is given by the Ginzburg-Landau equation typical of phase transition models.
We introduce the classical potentials describing second-order phase transitions (\cite{Fab})
$$
F(\vf)=\frac{1}{4}\vf^4-\frac{1}{2}\vf^2,\qquad\qquad G(\vf)=\frac{1}{2}\vf^2
$$
and assume the following equation
 \begin{equation}\label{GL}
	\tau\dot\vf=\frac{1}{\k^2}\Delta\vf-\t_c F'(\vf)-\t G'(\vf)-A(\vf,\t)I(\t_c-\t)\bH\cdot\bm
\end{equation}
where $\tau$, $\k^2$ are positive constants, $A$ is a generic function whose definition will be specified later, and $I$ is the unit step function, i.e.
$$I(x)=\begin{cases}
0 \qquad \text{ if\ }x<0\\
1 \qquad \text{ if\ } x\geq 0
\end{cases}$$
Notice that for large values of the temperature, $\t>\t_c$, last term of (\ref{GL}) vanishes and the function
$$
W(\vf)=\t_cF(\vf)+\t G(\vf)
$$
admits the minimum value $\vf=0$ which characterizes the paramagnetic phase.

The evolution equation for $\vf$ should preserve the defining condition $\vf\geq 0$ , which is not automatic in eq. (\ref{GL}). So this is a further constraint which has to be enforced, for example, in any numerical solution of the equation, and amounts to add a singular contribution in the potential function $F(\vf)$ such that $F(\vf)=+\infty$ for $\vf<0$.

Concerning the evolution of the direction of the magnetization, we assume that the unit versor $\bm$ satisfies the Landau-Lifshitz equation (\cite{LL1935})
\begin{equation}\label{LL2}
\vf^2\dot\bm=-\gamma\vf^2\bm\times\bH-\lambda\bm\times(\bm\times\bH),\qquad \gamma,\lambda>0.
\end{equation}
Notice that 
$$
\vf^2 |\dot\bm|=(\gamma^2\vf^4+\lambda^2)^{\frac12}|\bm\times\bH|.
$$
As a consequence, when $\vf$ approaches zero, the direction of $\bm$ moves toward the direction of $\bH$. Moreover, by multiplying (\ref{LL2}) by $\bm$ we obtain
$$
\vf^2 \dot\bm\cdot\bm=0,
$$
which is consistent with the condition $|\bm(x,t)|=1$ for any $t>0$, provided that $|\bm(x,0)|=1$.
 
Like other models of phase transitions (see \cite{Fab}), equation (\ref{GL}) can be interpreted as a balance law of the order structure. Indeed it can be written in the form
$$
k=\div\bp,
$$
where
\begin{eqnarray*}
k&=&\tau\dot\vf+\t_c F'(\vf)+\t G'(\vf)+A(\vf,\t)I(\t_c-\t)\bH\cdot\bm,\\
\bp&=&\frac{1}{\k^2}\nabla\vf.
\end{eqnarray*}
This formulation allows us to define the internal power related to the phase variable as
\begin{eqnarray*}
\cP_{\vf}&=&k\dot\vf+\bp\cdot\nabla\dot\vf\\
&=&\tau\dot\vf^2+\t_c \dot F(\vf)+\t \dot G(\vf)+A(\vf,\t)I(\t_c-\t)\dot \vf \bm\cdot\bH+\frac{1}{\k^2}\nabla\vf\cdot \nabla\dot\vf.
\end{eqnarray*}
From (\ref{B}) and (\ref{M}) we deduce the relation
$$
\dot\bB=\mu\dot\bH+\left(\frac{\partial\mu}{\partial\t}\dot\t+\frac{\partial\mu}{\partial\bH}\cdot\dot\bH\right)\bH+\left(\frac{\partial M}{\partial\vf}\dot\vf+\frac{\partial M}{\partial\t}\dot\t\right)\bm+M(\vf,\t)\dot\bm.
$$
Hence the electromagnetic power 
$$
\cP_{el}=\dot \bB \cdot\bH+\dot \bD\cdot \bE+\s \bE^2
$$
can be written as
\begin{eqnarray*}
\cP_{el}&=&\mu\dot\bH\cdot\bH+\left(\frac{\partial\mu}{\partial\t}\dot\t+\frac{\partial\mu}{\partial\bH}\cdot\dot\bH\right)\bH^2+\left(\frac{\partial M}{\partial\vf}\dot\vf+\frac{\partial M}{\partial\t}\dot\t\right)\bm\cdot\bH+M(\vf,\t)\dot\bm \cdot\bH\nonumber\\
&&+\e\dot\bE\cdot \bE+\s \bE^2.
\end{eqnarray*}
We denote by $e$ the internal energy and $h$ the thermal power. The first law of thermodynamics reads
\begin{equation}
\label{et}
\dot e=\cP_{el}+\cP_\vf+h.
\end{equation}
where $h$ satisfies the thermal balance law
\begin{equation}
\label{therm}
h=-\div \bq+r,
\end{equation}
and $\bq$, $r$ are respectively the heat flux and the heat source.

In order to prove the consistence of the model with the second law of thermodynamics we look for the constitutive relations for the entropy function $\eta$ and the heat flux $\bq$ that ensure the fulfillment of Clausius-Duhem inequality
$$
\dot\eta\geq -\div\left(\frac{\bq}{\t}\right)+\frac{r}{\t}.
$$
Thermal balance law (\ref{therm}) yields
$$
\t\dot\eta\geq \frac{\bq}{\t} \cdot\nabla\t +h.
$$
Hence, by introducing the free energy $\psi=e-\t\eta$, the previous inequality leads to
$$
\dot\psi+\eta\dot\t \leq \cP_{el}+\cP_{\vf}-\frac{\bq}{\t}\cdot \nabla\t.
$$
By substituting the expressions of the powers, we obtain
\begin{eqnarray*}
\dot\psi+\eta\dot\t  &\leq& \mu\dot\bH\cdot\bH+\left(\frac{\partial\mu}{\partial\t}\dot\t+\frac{\partial\mu}{\partial\bH}\cdot\dot\bH\right)\bH^2+\left(\frac{\partial M}{\partial\vf}\dot\vf+\frac{\partial M}{\partial\t}\dot\t\right)\bm\cdot\bH+M(\vf,\t)\dot\bm \cdot\bH\nonumber\\
&&+\e\dot\bE\cdot \bE+\s \bE^2+\tau\dot\vf^2+\t_c \dot F(\vf)+\t \dot G(\vf)+A(\vf,\t)I(\t_c-\t)\dot \vf \bm\cdot\bH\\
&&+\frac{1}{\k^2}\nabla\vf\cdot \nabla\dot\vf-\frac{\bq}{\t}\cdot \nabla\t.
\end{eqnarray*}
By means of (\ref{LL2}), we deduce
\begin{eqnarray*}
\dot\psi+\eta\dot\t  &\leq& \mu\dot\bH\cdot\bH+\left(\frac{\partial\mu}{\partial\t}\dot\t+\frac{\partial\mu}{\partial\bH}\cdot\dot\bH\right)\bH^2+\left(\frac{\partial M}{\partial\vf}\dot\vf+\frac{\partial M}{\partial\t}\dot\t\right)\bm\cdot\bH\nonumber\\
&&+\lambda M(\vf,\t)\vf^{-2} |\bm\times\bH|^2+\e\dot\bE\cdot \bE+\s \bE^2+\tau\dot\vf^2+\t_c \dot F(\vf)+\t \dot G(\vf)\\
&&+A(\vf,\t)I(\t_c-\t)\dot \vf \bm\cdot\bH+\frac{1}{\k^2}\nabla\vf\cdot \nabla\dot\vf-\frac{\bq}{\t}\cdot \nabla\t.
\end{eqnarray*}

We assume that the free energy $\psi$ depends on the variables $(\vf,\nabla\vf,\t,\bE,\bH)$, so that the previous
inequality yields
\begin{eqnarray}
&&\left[\frac{\partial \psi}{\partial\vf}-\left(A(\vf,\t) I(\t_c-\t)+\frac{\partial M}{\partial\vf}\right)\bm\cdot\bH-\t_c F'(\vf)-\t G'(\vf)\right]\dot\vf\nonumber
\\
&&+
\left[\frac{\partial \psi}{\partial\nabla\vf}-\frac{1}{\k^2}\nabla\vf\right]\cdot\nabla\dot\vf
+\left[\frac{\partial \psi}{\partial\t}+\eta-\frac{\partial\mu}{\partial\t}\bH^2-\frac{\partial M}{\partial\t}\bm\cdot\bH\right]\dot\t
+\left[\frac{\partial \psi}{\partial\bE}-\e\bE\right]\cdot\dot\bE
\nonumber
\\
&&+\left[\frac{\partial \psi}{\partial\bH}-\mu\bH-\frac{\partial\mu}{\partial\bH}\bH^2\right]\cdot\dot\bH
\nonumber
\\
&&\leq \lambda M(\vf,\t)\vf^{-2} |\bm\times\bH|^2
+\s \bE^2 +\tau \dot\vf^2-\frac{\bq}{\t}\cdot \nabla\t.
\label{CD}
\end{eqnarray}

The previous inequality is fulfilled if we choose the constitutive relations
\begin{eqnarray*}
&&A(\vf,\t) I(\t_c-\t)+\frac{\partial M}{\partial\vf}=0\\
&&\bq=-k_0(\t)\nabla\t\qquad k_0(\t)>0.
\end{eqnarray*}
Usual arguments of thermodynamics based on the arbitrariness of $(\dot\vf,\nabla\dot\vf,\dot\t,\dot\bE,\dot\bH)$ lead to
the following expressions of the free energy and entropy
\begin{eqnarray*}
\psi&=&\psi_0(\t)+ \frac{\e}{2}\bE^2+\mu\bH^2+\frac{1}{2\k^2}|\nabla\vf|^2-\int\mu(\bH,\t)\bH\cdot d\bH+\t_cF(\vf)+\t G(\vf)\\
\eta&=&-\frac{\partial\psi}{\partial\theta}+{\frac{\partial\mu}{\partial\t}\bH^2}+\frac{\partial M}{\partial\theta}\bm\cdot\bH\\
&=&-\psi_0'(\t)+\int\frac{\partial\mu}{\partial\t}\bH\cdot d\bH-G(\vf)+\frac{\partial M}{\partial\theta}\bm\cdot\bH.
\end{eqnarray*}

Substitution into (\ref{CD}) yields
$$
\lambda M(\vf,\t)\vf^{-2} |\bm\times\bH|^2
+\s \bE^2 +\tau \dot\vf^2+\frac{k_0(\t)}{\t} |\nabla\t|^2\geq 0,
$$
which guarantees that Clausius-Duhem inequality is satisfied.

The evolution equation for the temperature follows from the thermal balance law (\ref{therm}), by substituting the expression of $h$ deduced from the first law (\ref{et}).
Since the internal energy is written as
\begin{eqnarray*}
e&=&\psi+\t\eta=e_0(\t)+\frac{\e}{2} \bE^2+\mu\bH^2+\frac{1}{2\k^2}|\nabla\vf|^2\nonumber
\\
&&-\int\left(\mu-\t\frac{\partial\mu}{\partial\t}\right)\bH\cdot d\bH+ \t_c F(\vf)+\t \frac{\partial M}{\partial\theta}\bm\cdot\bH,
\end{eqnarray*}
where
$$
e_0(\t)=\psi_0(\t)-\t\psi_0'(\t),
$$
substitution into (\ref{et}) yields
\begin{eqnarray}
\label{h}
h&=&e_0'(\t)\dot \t- \t \dot G(\vf)+{\t\frac{\partial\mu}{\partial\t}\bH\cdot\dot\bH}+\t\frac{d}{dt}\left(\frac{\partial M}{\partial\t}\bm\cdot\bH\right)
+\t\dot\t\int\frac{\partial^2\mu}{\partial\t^2}\bH\cdot d\bH
\nonumber\\
&&-\tau \dot\vf^2-\s \bE^2-\lambda M(\vf,\t)\vf^{-2}|\bm\times\bH|^2.
\end{eqnarray}
Hence the temperature satisfies the equation
\begin{equation}\label{temp}
h=-\div[k_0(\t)\nabla\t]+r.
\end{equation}
In this model we assume the following constitutive equation
$$
A(\vf,\t)=\t-\t_c.
$$
This choice leads to a continuous temperature dependence  for the modulus of $\hat\bM$, namely
$$
M=\vf (\t-\t_c)_-,
$$
where the subscript $-$ denotes the negative part of a function, i.e.
$f_-=\max\{-f,0\}$.

Therefore the Ginzburg-Landau equation for the phase field reads
$$
	\tau\dot\vf=\frac{1}{\k^2}\Delta\vf-\t_c F'(\vf)-\t G'(\vf)+(\t-\t_c)_-\bH\cdot\bm.
$$

\section{One dimensional model}
In this section we will consider a one-dimensional model, obtained by assuming that the magnetic and electric fields have constant and orthogonal directions, say $y$, $z$, and that the components of the unknown fields depend only by the variable $x$, namely
$$
\bE=E(x)\bk,\qquad\qquad \bH=H(x)\bj.
$$
In the description of uniaxial ferromagnets, we modify the definition of the order parameter, by requiring that $\vf\neq 0$ in the ferromagnetic phase and $\vf=0$ in the paramagnetic state.
Therefore $\vf$ is allowed to take negative values and the vector $\bm$ is defined as
$$
\bm=sign(\vf)\bj.
$$
We assume the constitutive equations
\begin{equation}\label{MM}
M(\vf,\t)=|\vf|(\t-\t_c)_{-}.
\end{equation}

Therefore
$$
\hat\bM=\vf (\t-\t_c)_{-}\bj
$$
and equations (\ref{Max1})-(\ref{B}) imply
\begin{eqnarray*}
\e\dot E&=&\partial_x H-\s E\\
\mu\dot H +\frac{\partial\mu}{\partial\t}H\dot\t +\frac{\partial\mu}{\partial H}H\dot H&=&\partial_{x} E-\dot\vf(\t-\t_c)_{-} +\vf I(\t_c-\t)\dot\t.
\end{eqnarray*}
The evolution of the magnetization $\vf$ is governed by the Ginzburg-Landau equation 
\begin{equation}\label{vf}
	\tau\dot\vf=\frac{1}{k^2}\partial_{xx}\vf-\t_c F'(\vf)-\t G'(\vf)+(\t-\t_c)_-H.
\end{equation}
Finally, the evolution equation for the temperature is deduced by (\ref{h})  and (\ref{temp})
\begin{eqnarray}
 \left[e_0'(\t)+\t\int\frac{\partial^2\mu}{\partial\t^2}\bH\cdot d\bH\right]\dot \t- \t \dot G(\vf)+\t\frac{\partial\mu}{\partial\t}H\dot H-\t\frac{d}{dt}\left[I(\t_c-\t) \vf H\right]\nonumber\\
=\tau \dot\vf^2+\s E^2-\partial_x q+r.
\end{eqnarray}
A constitutive equation of $\mu(\theta,H)$ has to be given. For example, in the classical Landau model of ferromagnetism, the (total) magnetization $M$ as a function of the magnetic field and the temperature is given by
\begin{equation}
b_0M^3+a_0(\theta-\theta_c)M-H=0,
\end{equation}
from which it is obtained the permeability at $\theta>\theta_c$
\begin{equation}\label{GL-M}
\frac{\mu}{\mu_0}-1=\frac{1}{b_0 M(H,\theta)^2+a_0(\theta-\theta_c)}.
\end{equation}
For $H=0$ this equation provides the well known Curie-Weiss law for the susceptibility,
$$
\chi_0\propto\frac{1}{\theta-\theta_c}\qquad \theta>\theta_c.
$$
For $H\neq0$ the resultant permeability is a regular function of the temperature. In our model, it is required that, whatever the constitutive relation for $\mu$ is taken, the integral
$$
J(\theta,H)=\int_0^{\bH}\frac{\partial^2\mu}{\partial\t^2}\bH'\cdot d\bH'
$$ 
exists finite. We observe that for $H\rightarrow\infty$, $\partial^2\mu/\partial\t^2$ is expected to tend at zero for 
saturation reasons, while, in this respect, the permeability resulting from (\ref{GL-M}) is reasonable only at small fields. 
Moreover, the function $e_0'(\t)$ has to satisfy $e_0'(\t)+\t J(\t,H)>0$ for every $\t$ and $H$ to have a standard parabolic 
heat equation. 
\\
We see that equation (\ref{vf}) is able to recover the hysteresis diagram typical of the phenomenon of ferromagnetism.
 To this purpose, we will consider a spatially homogeneous material in isothermal conditions, with $\theta<\theta_c$. 
Then (\ref{vf}) reduces to 
\begin{equation}\label{vf_hom}
	\tau\dot\vf=-\t_c F'(\vf)-\t G'(\vf)-(\t-\t_c)H.
\end{equation}
Moreover, from (\ref{B}) and (\ref{M}) we obtain
\begin{equation}\label{B_hom}
B=\mu H-(\t-\t_c)\vf.
\end{equation}
Here we assume $\mu$ as approximately $H$-independent in the considered range of the magnetic field.
If $H$ is a known function of time, equations (\ref{vf_hom})-(\ref{B_hom}) allow us to obtain the $B-H$ diagram. 
In particular if $H=H_0\sin (\omega t)$, $t\in[0,T]$ and the initial condition is $\vf(0)=0$, we deduce the following 
 hysteresis diagrams for different values of the temperature.

\begin{figure}[ht]
\includegraphics{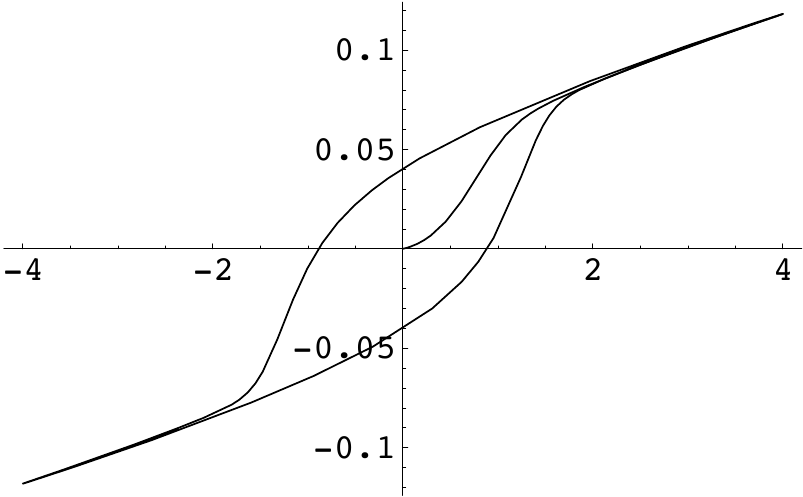}
\label{Fig_isteresi3}
\caption{$B-H$ diagram  with the numerical constants $
\t = 0.9,\quad\t_c = 1,\quad \omega=\pi,\quad\tau = 0.01,\quad H_0 = 4,\quad\mu = 0.01,\quad T = 2.5
$.}
\end{figure}

\begin{figure}[ht]
\includegraphics{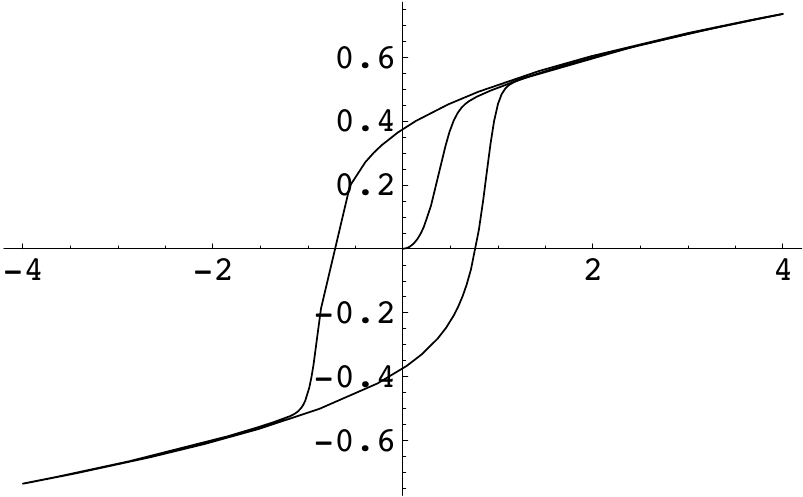}
\label{Fig_isteresi1}
\caption{$B-H$ diagram with the numerical constants $
\t = 0.5,\quad\t_c = 1,\quad \omega=\pi,\quad\tau = 0.01,\quad H_0 = 4,\quad\mu = 0.01,\quad T = 2.5
$.}
\end{figure}

\begin{figure}[ht]
\includegraphics{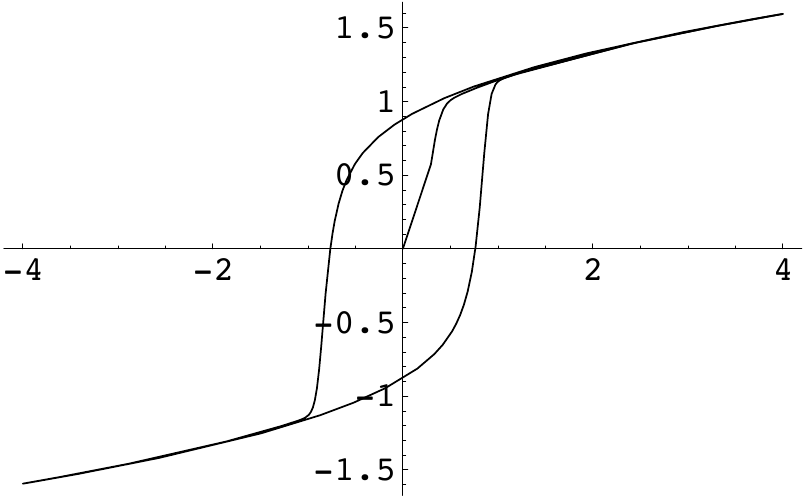}
\label{Fig_isteresi2}
\caption{$B-H$ diagram  with the numerical constants $
\t = 0.1,\quad\t_c = 1,\quad \omega=\pi,\quad\tau = 0.01,\quad H_0 = 4,\quad\mu = 0.01,\quad T = 2.5
$.}
\end{figure}

\end{document}